\documentclass[epj]{svjour}
\usepackage{latexsym}
\usepackage{graphics}
\usepackage{epstopdf}

\usepackage{amsmath}

\DeclareMathOperator{\sech}{sech}

\usepackage[numbers, square, comma, sort&compress]{natbib} 
\usepackage{color}

\usepackage{mathtools}

\newcommand{\difdisp}[2]{\frac{\mathrm{d}#1}{\mathrm{d}#2}} 
\newcommand{\difil}[2]{\mathrm{d}#1/\mathrm{d}#2} 
\newcommand{\pdifdisp}[2]{\frac{\partial #1}{\partial #2}}
\newcommand{\pdifil}[2]{\partial #1/\partial #2}

\newcommand{\e}[0]{\mathrm{e}}
\newcommand{\imi}[0]{\mathrm{i}} 

\DeclarePairedDelimiterX\braket[2]{\langle}{\rangle}{#1 \delimsize\vert #2}
\DeclarePairedDelimiter\abs{\lvert}{\rvert}%

 \DeclareFontFamily{U}{wncy}{}
    \DeclareFontShape{U}{wncy}{m}{n}{<->wncyr10}{}
    \DeclareSymbolFont{mcy}{U}{wncy}{m}{n}
    \DeclareMathSymbol{\Sh}{\mathord}{mcy}{"58}

\begin{document}
\title{Optical Lattice Trap for Kerr Solitons}
\author{Hossein Taheri, Andrey B. Matsko \and Lute Maleki%
\thanks{The authors acknowledge partial support from the Microsystems Technology Office of the Defense Advanced Research Projects Agency. H.T. thanks Kurt Wiesenfeld and Ali Eftekhar of the Georgia Institute of Technology for helpful discussions.
}%
}

\institute{OEwaves Inc., 465 North Halstead Street, Suite 140, Pasadena, CA 91107}
\date{Received: \today}

\abstract{
We show theoretically and numerically that dichromatic pumping of a nonlinear microresonator by two continuous wave coherent optical pumps creates an optical lattice trap that results in the localization of intra-cavity Kerr solitons with soliton positions defined by the beat frequency of the pumps. This phenomenon corresponds to the stabilization of the Kerr frequency comb repetition rate. The locking of the second pump, through adiabatic tuning of its frequency, to the comb generated by the first pump allows transitioning to single-soliton states, manipulating the position of Kerr solitons in the cavity, and tuning the frequency comb repetition rate within the locking range. It also explains soliton crystal formation in resonators supporting a dispersive wave emitted as a result of higher-order group velocity dispersion or avoided mode crossing. We show that dichromatic pumping by externally stabilized pumps can be utilized for stabilization of microresonator-based optical frequency combs when the comb span does not cover an octave or a significant fraction thereof and standard self-referencing techniques cannot be employed. Our findings have significant ramifications for high-precision applications of optical frequency combs in spectrally pure signal generation, metrology, and timekeeping.
}
\maketitle
\section{Introduction}
\label{intro}
Microresonator-based Kerr frequency combs, sometimes called ``microcombs'', are produced when optical resonators characterized with cubic (Kerr) nonlinearity are pumped with continuous wave (CW) light of sufficient power and certain frequency \cite{del2007monolithic, kippenberg2011microresonatorbased, savchenkov2016nanophotonics}. Coherent optical frequency combs correspond to stable states of oscillation in the nonlinear resonator. These states are represented by either hyperparametric oscillations  \cite{matsko2005hyperparametric} (Turing patterns \cite{Coillet2013Turing}) arising from modulational instability of the vacuum fluctuations, or by dissipative cavity Kerr solitons.  These latter objects are temporally localized structures in the microresonator \cite{matsko2011modelocked, chembo2013spatiotemporal, coen2013modeling, saha2013modelocking, herr2014temporal} that can be excited with nonadiabatic tuning of the pump laser driving the oscillation, or through modulating the laser phase or intensity \cite{taheri2015soliton, jang2015writing, lobanov2016harmonization}. Kerr frequency combs enable a number of applications such as high-speed optical telecommunication, high-spectral purity microwave and radio frequency (RF) signal generation, atomic optical clocks, metrology, astronomical spectrograph calibration, and search for exoplanets \cite{pfeifle2014coherent, pfeifle2015optimally, liang2015high, savchenkov2013clock, maleki2011rubidiumclock, papp2014clock, jost2015counting, kippenberg2011microresonatorbased}.

Stability and coherence of optical frequency combs is a key requirement for their applications. Self-referencing is one of the most commonly used techniques for stabilizing frequency combs \cite{bartels2009selfref}. This method requires the frequency span of the microcomb to cover a significant portion of an octave \cite{brasch2016photonic}. It is also possible to stabilize a microcomb by stabilizing the pump light together with the frequency spacing in the resonator producing the optical harmonics \cite{Savchenkov2013stabilization} or by locking two optical harmonics of the combs to two optical reference signals \cite{delhaye2008fullstab}. Comb stabilization in these schemes requires application of electronic control loops applied to various parameters of the pump laser.

\begin{figure}[tbp]
  \centering
  \includegraphics[width=8.5cm]{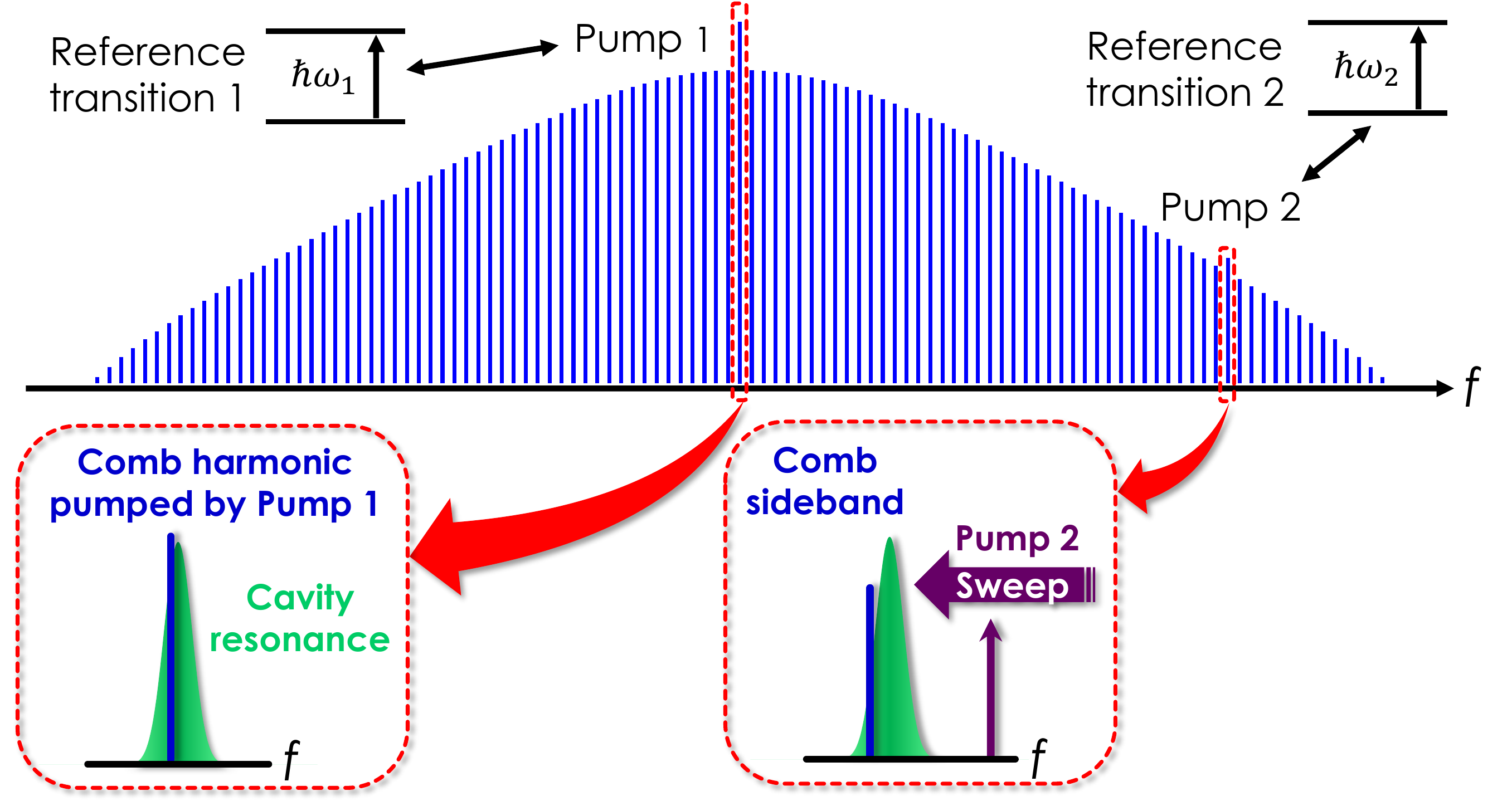}
\caption{ \small Dichromatic pumping of an optical frequency comb with pumps locked to reference transitions (e.g., to two different spectral lines in the same atom). By tuning the second pump into resonance after the generation of a frequency comb with the first pump, the second pump will lock to the microcomb via the frequency harmonic nearest to it. 
} \label{fig:BCScheme}
\end{figure}

In this work, we introduce another method for comb stabilization based on dichromatic\footnote{Some authors have preferred the word ``bichromatic'' \cite{strekalov2009generation, hansson2014bichromatically}. We use ``dichromatic'' because it combines the Greek prefix ``di'' (instead of the Latin prefix ``bi'') with ``chromatic'', a word of Greek origin. This choice creates a more lucid contrast with the commonly used adjective ``monochromatic''.} excitation of the microresonator, Fig.~\ref{fig:BCScheme}. We show, analytically and numerically, that driving the resonator producing a Kerr comb with a second pump creates an ``optical lattice'' and traps cavity solitons (similar to trapping of atoms in a standing wave \cite{letokhov1968narrowing, metcalf1999cooling}). Specifically, the application of the second light field together with the first pump results in a confining lattice that localizes two or multiple solitons in the resonator. These cavity solitons are separated by a temporal ``distance'' equal to multiples of the inverse of the beat frequency produced by the two pumps. As such, the application of the second laser stabilizes both the microcomb repetition rate and the optical frequency of its harmonics. Moreover, each of the pump lasers can be locked to a reference transition (e.g., the D$_1$ and D$_2$ transitions of rubidium, $^{87}$Rb, at 795 nm and 780 nm, if resonator dispersion is engineered and anomalous GVD is achieved in this wavelength range \cite{soltani2016arbitrary}) to improve the accuracy of the spectral position of comb harmonics \cite{savchenkov2013clock, papp2014clock}, thereby enabling  stabilization of a microcomb for ultra-pure RF and microwave signal generation, and optical atomic clock applications.

Dichromatic pumping and pump modulation were introduced earlier for microcomb generation and stabilization \cite{strekalov2009generation, papp2013parametric, taheri2014threshold, taheri2015soliton, hansson2014bichromatically, lobanov2016harmonization}, where the frequency spacing between the two pumps (for dual pumping) or between a pump and its adjacent harmonics (for pump modulation) determined the frequency comb repetition rate; the repetition rate was pinned to the difference between the seeding frequencies. In this work, in contrast, we show that repetition rate stabilization is possible even when a microcomb with a repetition rate corresponding to a single cavity FSR (free-spectral range) is pumped at frequencies separated by multiple FSRs.

We provide a generic mathematical formulation of light propagation in a dually-pumped Kerr-nonlinear microresonator, leading to a modified Lugiato-Lefever Equation (LLE) \cite{lugiato1987spatial}. In this formulation, power and detuning from the nearest resonance mode for both pumps are considered to be arbitrary. This formulation includes earlier models \cite{hansson2014bichromatically, okawachi2015dual} as a special case and results in understanding the formation and structure of temporarily ordered Kerr cavity soliton structures. Kerr solitons simultaneously confined in a microresonator with ideal second-order anomalous group velocity dispersion (GVD) interact weakly and can have nearly arbitrary relative positions. Their interaction becomes stronger as they are further packed when the number of solitons increases in the resonator. It is noted, however, that in actual microresonators, the solitons assume ordered positions around the resonator periphery \cite{lamb16stabilizing}. These ordered solitons, termed ``soliton crystals'', are analogous to ordered pulse structures observed in fiber lasers \cite{Amrani2010matter, Amrani2011crystals} and their generation in optical microresonators is attributed to the presence of avoided mode crossing (AMC) in microresonators \cite{savchenkov2012overmoded, herr2014mode}. Nonetheless, the role of AMC in the stablization of soliton crystals is not clear. In this paper we show that the ordering of the pulses originates from the dispersive wave (DW) generated due to  mode crossing \cite{matsko2016cherenkov}. The dispersive wave can be considered as a second coherent pump which locks the relative position of the pulses. A similar effect occurs where the presence of higher-order dispersion creates a coherent dispersive wave and leads to soliton crystal formation \cite{wang2014hod, milian2014soliton, taheri2016highorderdisp}.

This paper is organized as follows. The analytic formalism explaining formation of the optical lattice trap for solitons is presented in Section \ref{sec:formulation}. In Section \ref{sec:Xtal_DW}, examples of soliton crystal formation due to dispersive waves generated either because of resonator high-order dispersion or AMC are presented. A detailed study of dichromatic pumping and soliton trapping in the case of two independent coherent pumps interacting with a resonator characterized by anomalous second-order dispersion is presented in Section \ref{sec:Xtal_dichrom}. Section \ref{sec:conclusion} concludes the paper.

\section{Analytical Model}
\label{sec:formulation}

\subsection{Modified LLE}

We start from the set of coupled ordinary differential equations (ODEs) to derive a modified form of the LLE describing the generation of a Kerr frequency comb when two CW optical pumps are present. The reason for the derivation of the LLE from the set of coupled-wave equations is the natural way of introduction of an arbitrary number of pumps in this approach. We utilize the coupled-wave equations introduced in \cite{chembo2010modal} for the temporal evolution of the intra-cavity field harmonics $A_j$, where $|A_j|^2$ is the number of photons in the mode labeled $j$, and modify it by taking into account two independent pump fields with frequencies $\Omega_1$ and $\Omega_2$
\begin{align}\label{eq:bichrom:ODE}
\difdisp{A_j}{t} &= -\frac{\Delta\omega_j}{2} A_j \nonumber \\
& -\imi g \sum_{l, m, n} A_l A_m^* A_n \exp[\imi(\omega_l -\omega_m +\omega_n - \omega_j)t] \, \delta_{j_{lmn}j} \nonumber \\
& + \sqrt{\Delta\omega_{\mathrm{e},j}}\mathcal{F}_1 \exp[\imi(\Omega_{1}-\omega_j)t] \delta_{j_1 j} \nonumber \\
& + \sqrt{\Delta\omega_{\mathrm{e},j}}\mathcal{F}_2 \exp[\imi(\Omega_{2}-\omega_j)t] \delta_{j_2 j},
\end{align}
where $\Delta\omega_j$ is the full width at the half maximum (FWHM) of the corresponding optical mode, $j_{lmn} = l - m + n$ and $j$ are integers, $j_1$ and $j_2$ are the mode numbers for the two cavity resonances closest, respectively, to the pumps with frequencies $\Omega_1$ and $\Omega_2$, $\Delta\omega_{\mathrm{e},j}$ is the FWHM contributed by external coupling, and
\begin{equation}\label{eq:bichrom:LLEdrive}
\mathcal{F}_{k} = -\imi \sqrt{\frac{P_{\mathrm{in},k}}{\hbar\omega_{j_{k}}}} \e^{-\imi \phi_{\mathrm{in},k}}, \, k \in \{ 1, 2 \}.
\end{equation}
is the drive term related to the external pump with power $P_{\mathrm{in},j_{1,2}}$ and phase $\phi_{\mathrm{in},j_{1,2}}$ which pumps the resonance frequency $\omega_{j_{1,2}}$ of the resonator. The parameter $g$ is the four-wave mixing gain given by $g = n_2 c \hbar \omega_{j_0}^2 / (n_0^2 V_{j_0})$, in which $n_0$ and $n_2$ are the linear and nonlinear indices of refraction, $c$ is the vacuum speed of light, $\hbar$ is the reduced Planck constant, $\omega_{j_0}$ is the resonance frequency of the cavity mode pumped by the first pump, and $V_{j_0}$ is its effective nonlinear mode volume \cite{matsko2005hyperparametric, chembo2010modal}.

Frequencies of the resonator modes are identified by expanding modal frequencies around resonance labeled $j_0$ with frequency $\omega_{j_0}$,
\begin{equation}\label{eq:bichrom:disp}
\omega_j = \omega_{j_0} + D_1 (j-j_0) + \frac{1}{2!}D_2 (j-j_0)^2 + \frac{1}{3!} D_3 (j-j_0)^3 + \cdots .
\end{equation}
The spatiotemporal intra-cavity waveform can be found from
\begin{equation}\label{eq:bichrom:wvform}
A(\theta, t) = \sum_j A_j(t) \exp[\imi(\omega_j-\omega_{j_0})t-\imi(j-j_0)\theta].
\end{equation}
Using
\begin{equation}
\difdisp{A}{t} = \sum_j \left[\difdisp{A_j}{t}+\imi (\omega_j - \omega_{j_0}) A_j(t)\right] \e^{\imi(\omega_j-\omega_{j_0})t-\imi(j-j_0)\theta},
\end{equation}
and
\begin{equation}
\imi^n \frac{\partial^n A}{\partial \theta^n} = \sum_j (j-j_0)^n A_j(t) \e^{\imi(\omega_j-\omega_{j_0})t-\imi(j-j_0)\theta},
\end{equation}
together with Eqs.~(\ref{eq:bichrom:ODE}), (\ref{eq:bichrom:disp}),  and (\ref{eq:bichrom:wvform}), the spatiotemporal equation describing dichromatic comb generation can be found,
\begin{align} \label{eq:bichrom:PDE}
\frac{\partial A}{\partial t} = & \left( -\frac{\Delta \omega_{j_0}}{2} -\imi g |A|^2 + \sum_{n=1}^{N \ge 2} \imi^{n+1} \frac{D_n}{n!}\frac{\partial^n}{\partial\theta^n} \right) A \nonumber \\
& + \sqrt{\Delta\omega_{\mathrm{e},j_1}}\mathcal{F}_1 \e^{ \imi \left[ \sigma_1 t + D_{\mathrm{int}}(\eta = j_1-j_0) t + (j_1-j_0)(D_1 t - \theta) \right] } \nonumber \\
& + \sqrt{\Delta\omega_{\mathrm{e},j_2}}\mathcal{F}_2 \e^{ \imi \left[ \sigma_2 t + D_{\mathrm{int}}(\eta = j_2-j_0) t + (j_2-j_0)(D_1 t - \theta) \right] }.
\end{align}
Here, $N$ is an integer, $\sigma_k=\Omega_k - \omega_{j_k}$ (for $k \in \{1, 2 \}$) is the detuning of pump $k$ from its nearest cavity resonance, $\eta = j-j_0$ is the relative mode number (mode number with respect to mode labeled $j_0$), and $D_{\mathrm{int}}$ is the integrated or residual dispersion at $\omega_{j_0}$, and is a function of the relative mode number defined by
\begin{equation}
D_{\mathrm{int}}(\eta) = \frac{1}{2!} D_2 \eta^2 + \frac{1}{3!} D_3 \eta^3+\cdots .
\end{equation}
In deriving Eq.~(\ref{eq:bichrom:PDE}), the frequency dependence of the resonator FWHM has been ignored, i.e., we have assumed that $\Delta\omega_{j}=\Delta\omega_{j_0}$, for any $j$.

It is essential to remove the rapidly oscillating term $\exp[\imi(j_{1,2}-j_0)D_1 t]$ from Eq.~(\ref{eq:bichrom:PDE}), which is possible through moving to a rotating reference frame via the following change of variables
\begin{equation}\label{eq:bichrom:rotframe}
\theta - D_1 t = \theta', \qquad t = t'.
\end{equation}
This reference frame rotates with an angular velocity equal to $D_1$, i.e., FSR (in rad/s) at the spectral position of the first pump. Using
\begin{align}
\frac{\partial}{\partial\theta} &= \frac{\partial}{\partial \theta'} \\
\frac{\partial}{\partial t} &= -D_1 \frac{\partial}{\partial \theta'} + \frac{\partial}{\partial t'}.
\end{align}
we rewrite Eq.(\ref{eq:bichrom:PDE}) as
\begin{align} \label{eq:bichrom:PDErotframe}
\frac{\partial A}{\partial t} = & \left( -\frac{\Delta \omega_{j_0}}{2} -\imi g |A|^2 + \sum_{n=2}^{N \ge 2} \imi^{n+1} \frac{D_n}{n!}\frac{\partial^n}{\partial\theta^n} \right) A \nonumber \\
& + \sqrt{\Delta\omega_{\mathrm{e},j_1}} \mathcal{F}_1 \e^{ \imi \left[ \sigma_1 t + D_{\mathrm{int}}(\eta = j_1-j_0) t - (j_1-j_0) \theta \right] } \nonumber \\
& + \sqrt{\Delta\omega_{\mathrm{e},j_2}} \mathcal{F}_2 \e^{ \imi \left[ \sigma_2 t + D_{\mathrm{int}}(\eta = j_2-j_0) t - (j_2-j_0) \theta \right] }.
\end{align}
To simplify the notation, we have dropped the primes in the last step. Note that the $D_1 (\partial A/\partial\theta)$ term has disappeared and the forcing terms have taken a slightly simpler form. These simplifications bear significance from the perspective of numerical integration of this equation.

The derived equation coincided with the modified LLE derived earlier for symmetric dichromatic pumping with $N=2$ used in \cite{hansson2014bichromatically}. When the two pumps have the same power ($\mathcal{F}_1 = \mathcal{F}_2 = \mathcal{F}$) as well as the same detuning from their respective cavity resonances ($\sigma_1 = \sigma_2 = \sigma$), and are equally spaced (in units of FSR) from the center resonance labeled $j_0$ (i.e., $j_1 - j_0 = -(j_2 - j_0) = \eta_0 \neq 0$), Eq.~(\ref{eq:bichrom:PDErotframe}) simplifies to
\begin{align}
\frac{\partial A}{\partial t} = & \left( -\frac{\Delta \omega_{j_0}}{2} -\imi \tilde{\sigma} -\imi g |A|^2 + \sum_{n=2}^{N \ge 2} \imi^{n+1} \frac{D_n}{n!}\frac{\partial^n}{\partial\theta^n} \right) A \nonumber \\
& + 2 \sqrt{\Delta\omega_{\mathrm{e},j_0}} \mathcal{F} \cos(\eta_0 \theta).
\end{align}
where $\tilde{\sigma} = \sigma + D_{\mathrm{int}}(\eta_0)$. To arrive at the above equation, the detuning has been absorbed in the field envelope, i.e., $A \to A \exp(-\imi \sigma t)$.

In what follows we consider the case when one of the pumps is at the center frequency $\omega_{j_0}$ (say, $j_2 = j_0$, $\eta_2 = 0$, $\sigma_2 = \sigma_0$, and $\mathcal{F}_2 = \mathcal{F}_0$), and the other pump is at another resonance ($\eta_1 \ne 0$).  Then  Eq.~(\ref{eq:bichrom:PDErotframe}) can be written in the form
\begin{align}\label{eq:bichrom:P1Center}
\frac{\partial A}{\partial t} = & \left( -\frac{\Delta \omega_{j_0}}{2} -\imi \sigma_0 -\imi g |A|^2 + \sum_{n=2}^{N \ge 2} \imi^{n+1} \frac{D_n}{n!}\frac{\partial^n}{\partial\theta^n} \right) A \nonumber \\
& + \sqrt{\Delta\omega_{\mathrm{e},j_0}} \left[ \mathcal{F}_0 + \mathcal{F}_1 \e^{ \imi (\sigma_1-\sigma_0)t + \imi D_\mathrm{int}(\eta = \eta_1) t - \imi \eta_1 \theta } \right].
\end{align}
Here, again, we have absorbed the detuning (of one of the pumps) in the field envelope, i.e., $A \to A \exp(-\imi \sigma_0 t)$.

\subsection{The force applied to a soliton}
\label{sec:solforce}

We now consider the influence of the second pump on a Kerr cavity soliton focusing on Eq.~(\ref{eq:bichrom:P1Center}). This case is practically significant because it will allow us to understand the effect of the second pump, in terms of stability and coherence, on a pulse initially generated in a microresonator with the first pump. We utilize the non-dimensional form of Eq.~(\ref{eq:bichrom:P1Center}) which reads
\begin{align} \label{eq:bichrom:LLE_nondim}
\frac{1}{\gamma} \frac{\partial \psi}{\partial t} &= \left[ -1 + \imi \left( \frac{\sigma_0}{\gamma} \right) + \imi |\psi|^2 - \sum_{n=2}^{N \ge 2} (-\imi)^{n+1} \frac{d_n}{n!}\frac{\partial^n}{\partial\theta^n} \right] \psi \nonumber \\
& + F(\theta, t),
\end{align}
where $\gamma = \Delta \omega_{j_0} / 2$ is the half-width at half-maximum (HWHM) of the resonance with frequency $\omega_{j_0}$, and
\begin{equation}\label{eq:bichrom:drive}
F(\theta, t) = F_0 + F_1 \exp \left[ \imi \eta_1 \theta - \imi (\sigma_1 -  \sigma_0) t - \imi D_\mathrm{int} (\eta_1) t \right].
\end{equation}
In Eqs.~(\ref{eq:bichrom:LLE_nondim}) and (\ref{eq:bichrom:drive}), the following normalizations have been used:
\begin{align}
d_n & =\frac{-D_n}{\gamma} , \\
\psi &= \frac{A^*}{\sqrt{ \gamma / g }}, \\
F_k &= \frac{\sqrt{\Delta \omega_{\mathrm{e},j_k}}}{\gamma} \frac{\mathcal{F}_k^*}{\sqrt{ \gamma / g }} \nonumber \\
&= \imi \sqrt{\frac{g \Delta \omega_{\mathrm{e},j_k}}{\gamma^{3}} \frac{P_{\mathrm{in},k}}{\hbar \omega_{j_k}}} \e^{\imi \phi_{\mathrm{in},k}}, \, k \in \{ 0, 1 \} \label{eq:bichrom:Fk}.
\end{align}
It is noteworthy that the normalized power inside the resonator is related to the physical intra-cavity power $P_\mathrm{r}$ through $\abs{\psi}^2 = g P_\mathrm{r} T_\mathrm{R} / (\gamma \hbar \omega_{j_0})$, where $T_\mathrm{R}$ is the cavity round-trip time. In physical terms, the $n$-th order dispersion coefficient $D_n$ has been normalized to the resonance bandwidth, and the intra-cavity field envelope and driving power have been normalized to sideband generation threshold $\sqrt{\gamma / g}$ \cite{matsko2005hyperparametric, chembo2010modal}. Note that $\tau = \gamma t$ can be considered a normalized time variable, i.e., laboratory time normalized to twice the photon lifetime $\tau_\mathrm{ph} = 1/\Delta \omega_{j_0}$. Equation (\ref{eq:bichrom:LLE_nondim}) reduced to the \emph{standard} LLE \cite{lugiato1987spatial, chembo2013spatiotemporal} when  higher-order dispersion terms are neglected (i.e., $d_n = 0$ for $n \ge 3$) and the second pump is non-existent or couples no power to the resonator (i.e., $F_1$).

Let us introduce the pulse momentum parameter \cite{barashenkov2004traveling} defined in terms of the intra-cavity field $\psi(\theta, t)$ by
\begin{equation}\label{eq:bichrom:P}
\mathcal{P} = \frac{1}{2} \int_{0}^{2\pi} \mathrm{d} \theta \, \psi^* \left( -\imi \frac{\partial}{\partial \theta} \right) \psi + \mathrm{c.c.} \, .
\end{equation}
To find the equation of motion for the momentum $\mathcal{P}$, we take the derivative of Eq.~(\ref{eq:bichrom:P}) with respect to the normalized time $\tau$ and get
\begin{align}\label{eq:bichrom:deriv1}
\difdisp{\mathcal{P}}{\tau} &= \frac{1}{2} \int_{0}^{2\pi} \mathrm{d} \theta \,  \left[ \pdifdisp{\psi^*}{\tau} \left( -\imi \pdifdisp{}{\theta} \right) \psi + \psi^* \left( -\imi \frac{\partial^2}{\partial\theta \partial \tau} \right) \psi \right] \nonumber \\
& + \mathrm{c.c.} \, .
\end{align}
The second term in the integrand in Eq.~(\ref{eq:bichrom:deriv1}) can be simplified using integration by parts and employing the periodicity of $\psi(\theta, t)$, so that
\begin{equation}
\difdisp{\mathcal{P}}{\tau} = \int_{0}^{2\pi} \mathrm{d} \theta \, \left[ \pdifdisp{\psi^*}{\tau} \left( -\imi \pdifdisp{}{\theta} \right) \psi  \right] + \mathrm{c.c.},
\end{equation}
which corresponds to the equation for the pulse momentum derived using variational approach \cite{vlasov1971averaged,Santhanam2003moments}.

Replacement of $\pdifil{\psi}{\tau}$ by the right-hand side (RHS) of Eq.~(\ref{eq:bichrom:LLE_nondim}) results in
\begin{align}\label{eq:bichrom:deriv2}
\difdisp{\mathcal{P}}{\tau} &= - 2 \mathcal{P} - \sum_{n = 2}^{N \ge 2} \frac{d_n}{n!} \int_{0}^{2\pi} \mathrm{d} \theta \, \left[ \imi^n \psi \frac{\partial^{n+1}}{\partial \theta^{n+1}} \psi^* + \mathrm{c.c.} \right] \nonumber \\
& + \int_{0}^{2\pi} \mathrm{d} \theta \, \left[ \psi^* \left( -\imi \frac{\partial}{\partial \theta} \right) F + \mathrm{c.c.} \right].
\end{align}
The second term on the RHS of Eq.~(\ref{eq:bichrom:deriv2}) can be further simplified using the discrete Fourier transform expansion
\begin{equation}
\psi(\theta, t) = \sum_{m=-\infty}^{+\infty} \tilde{a}_m (t) \, \e^{\imi m \theta},
\end{equation}
where $m = j - j_0$ is the mode number relative to the pumped mode and $\tilde{a}_m$ is the complex-valued amplitude of the comb tooth labeled $m$, normalized to the sideband generation threshold; cf. Eq.~(\ref{eq:bichrom:wvform}). Recalling that $ \int_{0}^{2\pi} \mathrm{d} \theta \, \exp [ \imi (m - m') ] = 2 \pi \delta_{m' m}$, with $\delta_{m' m}$ denoting the Kronecker's delta, we find that
\begin{equation}
\int_{0}^{2\pi} \mathrm{d} \theta \, \left[ \imi^n \psi \frac{\partial^{n+1}}{\partial \theta^{n+1}} \psi^* \right] = -2 \pi \imi \sum_m m^{n+1} \abs{\tilde{a}_m}^2.
\end{equation}
This term is pure imaginary and therefore the second term on the RHS of Eq.~(\ref{eq:bichrom:deriv2}) is zero,  so that the equation transforms to
\begin{equation} \label{eq:bichrom:dPdt}
\frac{1}{\gamma} \difdisp{\mathcal{P}}{t} = -2 \mathcal{P} + \int_{0}^{2\pi} \mathrm{d} \theta \, \left[ \psi^* \left( -\imi \frac{\partial}{\partial \theta} \right) F + \mathrm{c.c.} \right].
\end{equation}
This expression shows that when the drive term $F$ does not depend on $\theta$, the equation of motion of the momentum parameter is $\difil{\mathcal{P}}{t} = -2 \gamma \mathcal{P}$ and, hence, the steady-state momentum is zero. However, when $F$ is a non-trivial function of $\theta$, as in the case of the dichromatic pump of Eq.~(\ref{eq:bichrom:drive}), the rotational symmetry of the system breaks down and an extra force appears. Equation (\ref{eq:bichrom:dPdt}), therefore, suggests that, if a soliton is generated in the cavity, the addition of the second pump applies a force on it. Since this force is a function of both the drive term $F(\theta, t)$ and the intra-cavity field $\psi(\theta, t)$, it may vanish for a particular waveform profile $\psi$ or a certain positioning of intra-cavity pulse with respect to the drive term $F$.

To understand how a Kerr soliton is affected by the force attributed to the second pump, we recall that the standard LLE (found from Eq.~(\ref{eq:bichrom:LLE_nondim}) for $N = 2$ and $F_1 = 0$) has an approximate solution in the form of a soliton atop a CW background \cite{matsko2011modelocked, matsko2013timing, herr2014temporal} given by
\begin{equation}
\psi(\theta) = \tilde{C}_1 + \tilde{C}_2 \sech ( B \theta ),
\end{equation}
where $\tilde{C}_1$ is the CW background found from
\begin{equation}
\tilde{C}_1 = \frac{\gamma F_0}{\sigma_0} \left( \frac{\gamma}{\sigma_0} - \imi \right) = C_1 \exp (\imi \phi_{C_1}),
\end{equation}
while
\begin{equation}
\tilde{C}_2 = \frac{4 \sigma_0}{\pi \gamma F_0} + \imi \sqrt{\frac{2 \sigma_0}{\gamma} - \left( \frac{4 \sigma_0}{\pi \gamma F_0} \right)^2} = C_2 \exp (\imi \phi_{C_2}),
\end{equation}
and
\begin{equation}
B = \sqrt{\frac{2\sigma_0}{\gamma}}.
\end{equation}
The normalized amplitude of the pump $F_0$ is found from Eq.~(\ref{eq:bichrom:Fk}) for $k = 0$, and $\sigma_0$ is detuning between the pump and its nearest cavity resonant mode.

Because of periodic boundary conditions, the sech-like solution is strictly valid far from the domain boundaries 0 and $2\pi$. This assumption will not be restrictive here since the soliton width is much narrower than the resonator periphery and $\eta_1 > 1 $ guarantees that studying the force over a range of $2 \pi / \eta_1$ is sufficient for understanding the effect of the second pump.

Substituting the soliton solution into Eq.~(\ref{eq:bichrom:dPdt}), we get an expression for the force
\begin{equation}\label{eq:bichrom:Pforce}
\tilde{C}_3 \int_{0}^{2\pi} \mathrm{d} \theta \, \psi^* \exp(\imi \eta_1 \theta) + \mathrm{c.c.},
\end{equation}
with $\tilde{C}_3 = \eta_1 F_1 \exp \left[ - \imi (\sigma_1 - \sigma_0) t - \imi D_\mathrm{int} (\eta_1) t \right]$. The CW background does not exert a force on a soliton, because
\begin{equation}
\tilde{C}_3 \tilde{C}_1^* \int_{0}^{2\pi} \mathrm{d} \theta \, \e^{\imi \eta_1 \theta} = 0.
\end{equation}
As for the force due to the influence of the second pump, the following integral needs to be evaluated
\begin{equation}
\tilde{C}_3 \tilde{C}_2^* \int_{0}^{2\pi} \mathrm{d} \theta \, \sech \left[ B(\theta - \theta_0) \right] \e^{\imi \eta_1 \theta},
\end{equation}
where $\theta_0$ is the pulse peak position. This integral can be expressed in terms of hypergeometric functions. For a narrow soliton pulse ($B \gg 1$), it can be approximated by
\begin{equation}\label{eq:bichrom:s1}
\tilde{C}_3 \tilde{C}_2^* \e^{\imi \eta_1 \theta_0} \int_{0}^{2\pi} \mathrm{d} \theta \, \sech \left[ B(\theta - \theta_0) \right] = \tilde{C}_3 \tilde{C}_2^* S \e^{\imi \eta_1 \theta_0},
\end{equation}
where $S = \int_{0}^{2\pi} \mathrm{d} \theta \, \sech \left[ B(\theta - \theta_0) \right]$ is the area under the $\sech(B \theta )$ curve. Equation (\ref{eq:bichrom:s1}) shows that, while we have used a sech soliton pulse, the present analysis holds true for any pulse profile because only the pulse area appears in the final expression. The momentum equation of motion, Eq.~(\ref{eq:bichrom:dPdt}), therefore, takes the form
\begin{equation}\label{eq:bichrom:dPdtComplete}
\frac{1}{\gamma} \difdisp{\mathcal{P}}{t} = -2 \mathcal{P} + 2 \eta_1 F_1 C_2 S \cos \left( \eta_1 \theta_0 - \tilde{\omega} t - \phi_{C_2} \right).
\end{equation}
where $\tilde{\omega} =  \sigma_1 - \sigma_0 + D_\mathrm{int}(\eta_1)$.

Equation (\ref{eq:bichrom:dPdtComplete}) suggests that the second pump can lead to a force. The force is time independent if the pump-resonance detuning values are chosen so that $ \tilde{\omega} = 0 $. The force is zero if
\begin{equation}\label{eq:bichrom:SSpositions}
\theta_{0}^{\mathrm{SS}} = \frac{2 k' \pi \pm \pi /2 + \phi_{C_2} }{\eta_1}
\end{equation}
where $k'$ is an integer. This result establishes that the presence of the properly tuned second pump creates a lattice around the resonator where the solitons are trapped in steady state. We note that the second pump in this study also emulates, in a simplified picture, the appearance in the Kerr comb power spectrum of a strong dispersive wave peak (which may appear due to resonator higher-order chromatic modal dispersion) or a strong comb tooth resulting from resonator avoided mode crossing. The foregoing discussion also shows that the presence of a strong comb tooth (apart from the pumped mode) in the microcomb power spectrum suppresses timing jitter \cite{matsko2013timing}, in the sense that any disturbance which tends to advance or delay the soliton will be opposed and the soliton will be pushed back to it ``site'' in the lattice created by the beating of the pumped mode and strong comb sideband.

\section{Soliton Crystals in the Presence of a Dispersive Wave}
\label{sec:Xtal_DW}

When the Kerr cavity soliton becomes phase-matched with one or more dispersive waves (DWs) in the resonator, sharply-peaked spectral features appear in the power spectrum envelope of the microcomb. The phase matching is supported either by the frequency-dependent GVD of the resonator modes or because of the interaction among the mode families of the resonator. For DW emission to occur, the group velocity of the optical pulse should match the phase velocity of the DW and the power spectrum of the pulse should overlap with the dispersive wave. It has been shown that under certain conditions the DW can become phase-locked to the harmonics of the Kerr frequency comb \cite{akhmediev1995cherenkov, milian2014soliton, taheri2016highorderdisp, herr2014mode, matsko2016cherenkov, vahala2016dw}. When this condition is fulfilled, the dispersive wave can be considered as a harmonic of the microcomb and a coherent part thereof.

We noted earlier that one of the effects of the dispersive wave on the Kerr frequency comb is related to the formation of so called soliton crystals. The beating of the DW with the coherent optical pump modulates the CW background of the spatiotemporal waveform of the intra-cavity pulse and, according to Section \ref{sec:formulation}, results in the creation of a grid of potentials with preferential pulse positions around the resonator circumference, namely, an optical lattice trap. The temporal spacing between the pulses trapped at the lattice sites around the resonator is determined by the frequency separation of the pump and the dispersive, i.e., their beating frequency. In this Section, we corroborate this prediction emphasized by Eq.~(\ref{eq:bichrom:SSpositions}) through numerical simulations based on both the integration of the nonlinear coupled-wave equations \cite{chembo2010modal} and the split-step integration of the LLE \cite{coen2013modeling, chembo2013spatiotemporal}. Both approaches give similar results.

\subsection{Dispersive wave due to avoided mode crossing}

Linear interaction of different resonator mode families in overmoded resonators or resulting from the resonator imperfections leads to abrupt disruptions in the spectrum of microcomb-hosting modes. The resulting avoided mode crossing (AMC), or mode anti-crossing, in the modal spectrum can be implemented directly in the coupled-wave equations of Kerr comb generation, or through a two-parameter fit,
\begin{align}\label{eq:bichrom:MCdisp}
\omega_j &= \omega_{j_0} + D_1 (j-j_0) + \frac{1}{2!}D_2 (j-j_0)^2 \nonumber \\
& \, + \frac{0.5 a}{j - j_0 - b - 0.5},
\end{align}
for the integration of the LLE, in which $a$ is a measure of the magnitude and $b$ specifies the spacing with respect to the pumped mode $j_0$ of the mode crossing \cite{herr2014mode}.

\begin{figure}[tbp]
  \centering
  \includegraphics[width=8.5cm]{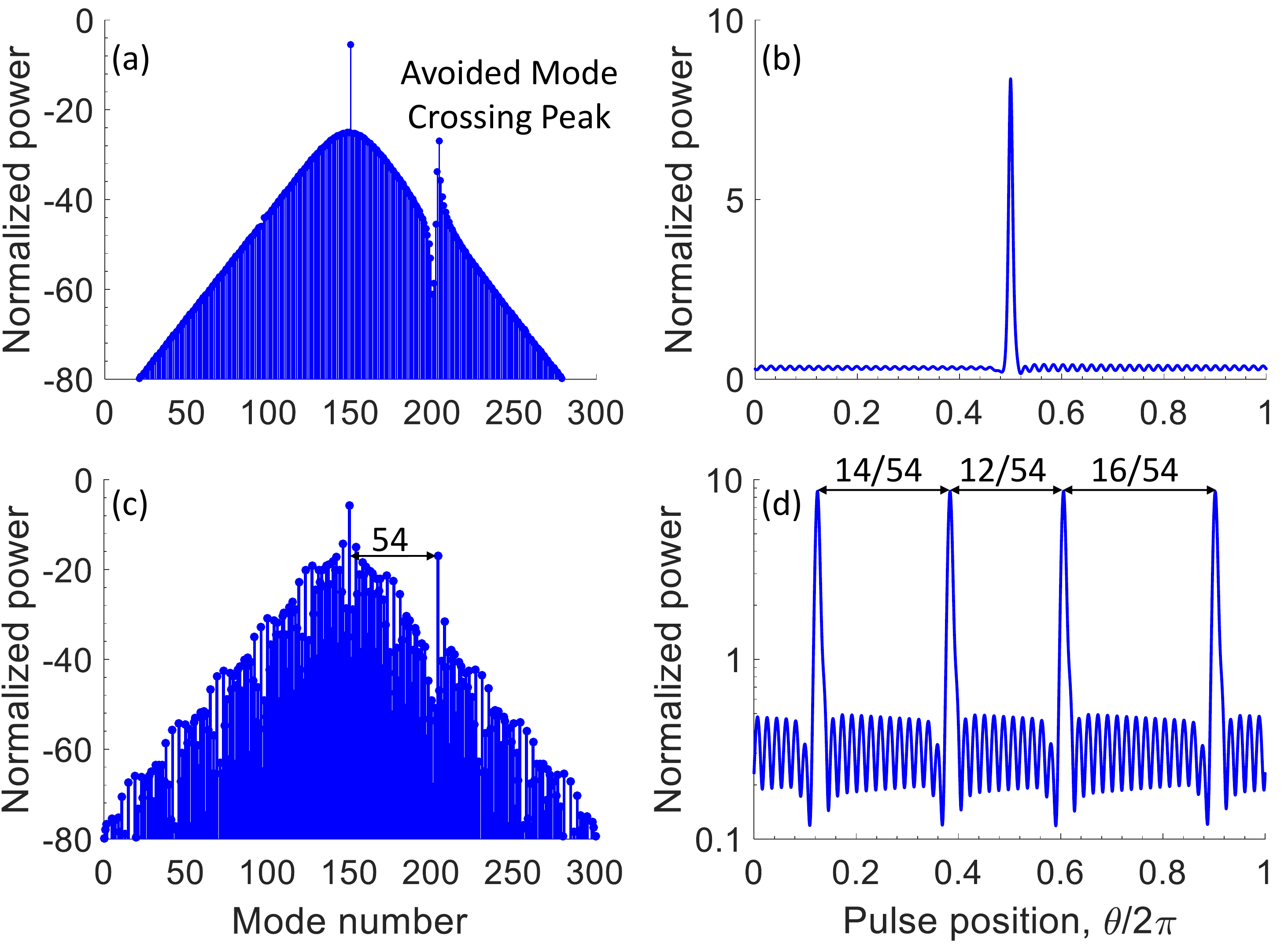}
\caption{ \small Kerr frequency comb in the presence of microresonator avoided mode crossing characterized by $D_\mathrm{int}(j-j_0)/\gamma=0.005(j - j_0)^2 + 0.5 a / (j - j_0 - b - 0.5)$, $a = 100$, $b = 50$, i.e., avoided mode crossing at $j - j_0 = 50$. (a) Steady-state power spectrum of a cavity soliton, with a strong comb sideband caused by mode crossing. The strong AMC peak appears at $j - j_0 = 54$. (b) The pulse waveform corresponding to the spectrum shown in (a). The signature of the AMC in the spatiotemporal waveform is the modulated pulse pedestal. (c) The AMC peak appears at the same spectral position when multiple solitons are confined in the resonator within one round-trip time. (d) The intra-cavity waveform for (c), where the separation between the pulses generated in the resonator is determined by the beat frequency of the pump and the AMC signature. To better visualized the lattice created in the resonator, the intra-cavity waveform is shown in logarithmic scale. A total number of 512 modes are used in the split-step Fourier transform integration of the LLE, where the continuous wave pump $|F_0|^2=3.85$ is applied to the central mode with detuning $\sigma_0 = -2.7 \gamma$.
} \label{fig:figMC}
\end{figure}

An example is shown in Fig.~\ref{fig:figMC}, where the comb harmonic resulting from the distortion of the resonator spectrum due to the AMC, Fig.~\ref{fig:figMC}(a), interacts with the pump light and modulates the CW pedestal of the cavity soliton as seen in Fig.~\ref{fig:figMC}(b). We found that when multiple solitons are confined within the resonator, they reside at preferential points around the resonator and therefore their separation at steady-state is quantized. Panel (d) of Fig.~\ref{fig:figMC} illustrates this notion; the temporal spacing step is determined by the spectral separation (beating frequency) between the DW and the pump, which in this examples equals 54 FSRs of the resonator at the spectral position of the pump, Fig.~\ref{fig:figMC}(c). This observation is in excellent agreement with Eq.~(\ref{eq:bichrom:SSpositions}).

\subsection{Dispersive wave due to higher-order modal dispersion}

High-order microresonator modal dispersion can lead to the emission of part of the soliton power into a dispersive wave and the generation of a few strong comb sidebands in the microcomb power spectrum \cite{milian2014soliton, taheri2016highorderdisp}. While the emission frequency can be found via a mathematically rigorous method \cite{milian2014soliton}, an intuitive approach can be taken to determine the approximate spectral position of the dispersive wave $j_\mathrm{DW}$ with respect to the cavity soliton center frequency $j_\mathrm{S}$ according to
\begin{equation}\label{eq:bichrom:DWphasematch}
\sum_{n = 1}^{N>2} \frac{D_n}{n!}(j_\mathrm{DW}-j_\mathrm{S})^n + g P_\mathrm{S} = 0,
\end{equation}
where $P_\mathrm{S} = \abs{\tilde{C}_2}^2$ is the soliton peak power, with $\tilde{C}_2$ defined in Section \ref{sec:solforce}. This expression is the phase matching condition for DW emission, which is similar to that used in optical fibers \cite{agrawal2013NLfiber} and emphasizes the physical origin of microcomb Cherenkov radiation based on the phase matching of the soliton with the DW \cite{akhmediev1995cherenkov}. Much like the case of the mode anti-crossing discussed earlier in this Section, the beating of the DW and the pump modulates the CW background of the intra-cavity waveform and creates an optical lattice with steps determined by $\abs{j_\mathrm{DW}-j_0}$.

\begin{figure}[tb]
  \centering
  \includegraphics[width=8.5cm]{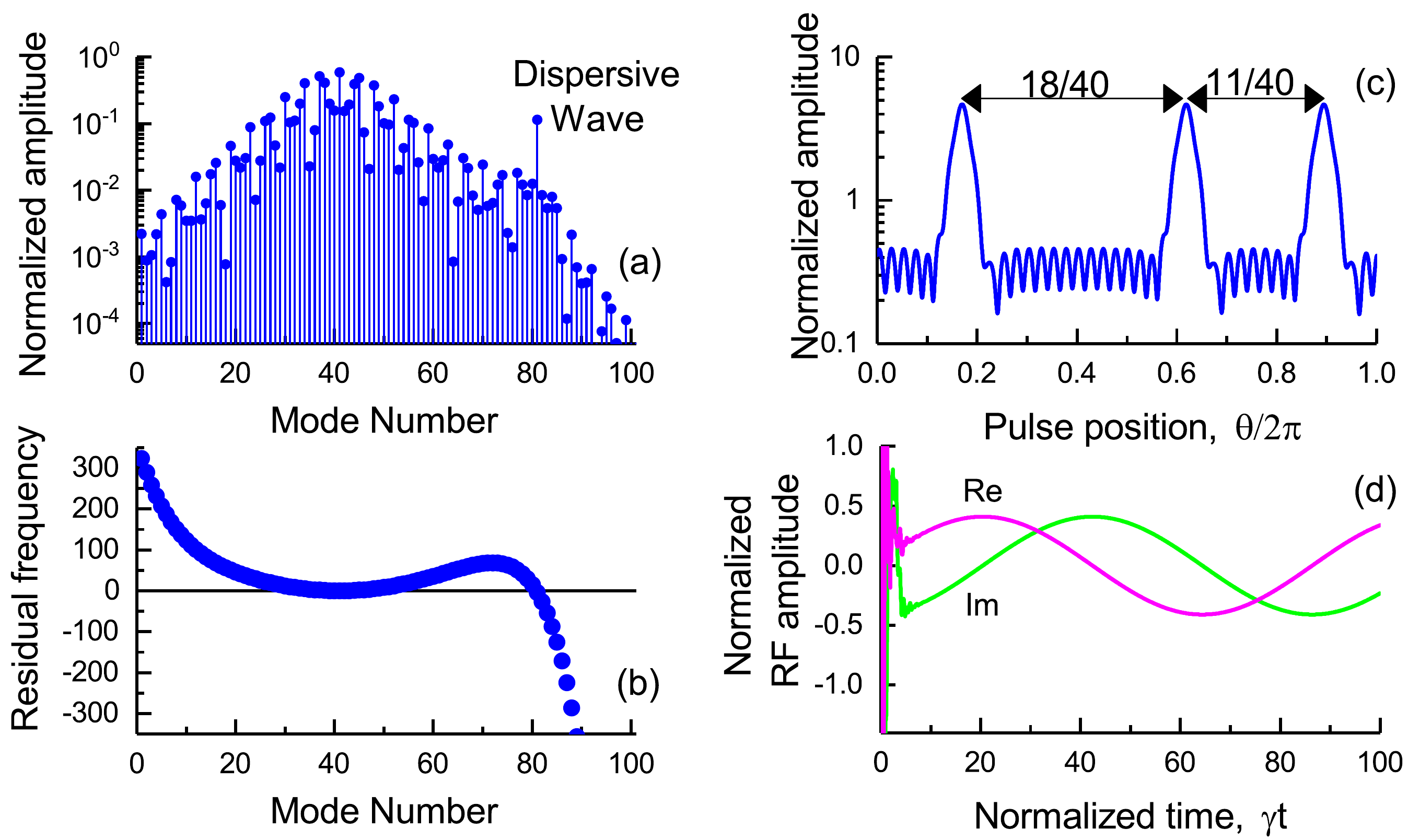}
\caption{ \small Kerr frequency comb with a coherent dispersive wave, (a), generated for specific residual frequency of the resonator $D_\mathrm{int}(j-j_0)/\gamma=0.1(j - j_0)^2 - 10^{-9}(j - j_0)^7$, depicted in (b). The continuous wave pump $|F_0|^2=10$ is applied to mode $41$ with detuning $\sigma_0 = -10 \gamma$, and the dispersive wave is generated in mode 81 of the 101 modes used in the numerical simulation. The position of the pulses generated in the resonator is determined by the beat frequency between the pump and the dispersive wave, (c). The presence of the dispersive wave shifts the frequency of the RF signal generated by the frequency comb and found by demodulating the comb on a photodiode. The frequency shift corresponds to the time dependent modulation of the real and imaginary parts of the RF signal, as illustrated in panel (d).
} \label{fig:DW}
\end{figure}

We present an example in Fig.~\ref{fig:DW}. The power spectrum of a multi-soliton micrcomb demonstrates a DW where Eq.~(\ref{eq:bichrom:DWphasematch}) is satisfied. The position of the soliton peaks in the intra-cavity waveform are determined by the beatnote of the DW and the pump (here, $j_\mathrm{DW}-j_0 = 40$), again confirming Eq.~(\ref{eq:bichrom:SSpositions}).

To show that the dispersive wave impacts the frequency comb repetition rate, we depict the RF signal which can be measured experimentally by demodulating the frequency comb on a photodiode (see Fig.~\ref{fig:DW}(d)). The slow complex amplitude of the RF signal is given by
\begin{equation}
i_{\mathrm{RF}} \sim \sum_j A_{j-1}^*A_j \e^{\imi D_1 t},
\end{equation}
where $D_1 = \omega_{\mathrm{FSR}}(j_0) = (\omega_{j_0+1}+\omega_{j_0-1})/2$ is the free spectral range of the resonator at the pump frequency; see Eq.~(\ref{eq:bichrom:disp}). For the case of the spectrum characterized with ideal anomalous GVD, the RF signal frequency is the same as the pumped mode FSR $\omega_{\mathrm{FSR}}(j_0)$ and the RF signal does not depend on time. However, the presence of higher-order dispersion as well as the dispersive wave changes the microcomb repetition rate. The resultant RF frequency becomes different from the local FSR and the real and imaginary amplitudes of the RF signal become time dependent.

\begin{figure}[th]
  \centering
  \includegraphics[width=8.5cm]{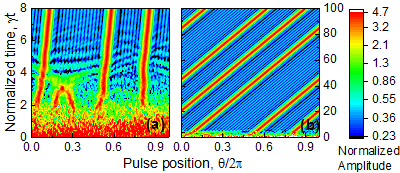}
\caption{ \small Formation of an optical crystal consisting of three solitons. The pulse amplitude $|A|$ is shown by the color density (logarithmic scale). The continuous wave pump is applied to the 41$^{\mathrm{st}}$ simulated mode. The parameters of the system are described in the caption of Fig.~(\ref{fig:DW}). Multiple pulses generated starting from random initial conditions (a), propagate stably for hundreds of cavity ring-down times parallel to the temporal propagation direction of the background modulation, i.e., trapped in the lattice generated because of the beating between the pumped mode and the dispersive wave, (b).
} \label{fig:DWXtal}
\end{figure}

Figure~\ref{fig:DWXtal} illustrates the formation of the soliton crystal depicted in Fig.~\ref{fig:DW}. The resultant pulses propagate synchronously along with the modulated background of the intra-cavity field. The time scale of the forming of the the pulses and the beatnote is approximately the same, as can be seen from Fig.~\ref{fig:DWXtal}(a).

\section{Soliton crystals induced by a dichromatic pump}
\label{sec:Xtal_dichrom}

\begin{figure}[tbp]
  \centering
  \includegraphics[width=0.5\textwidth]{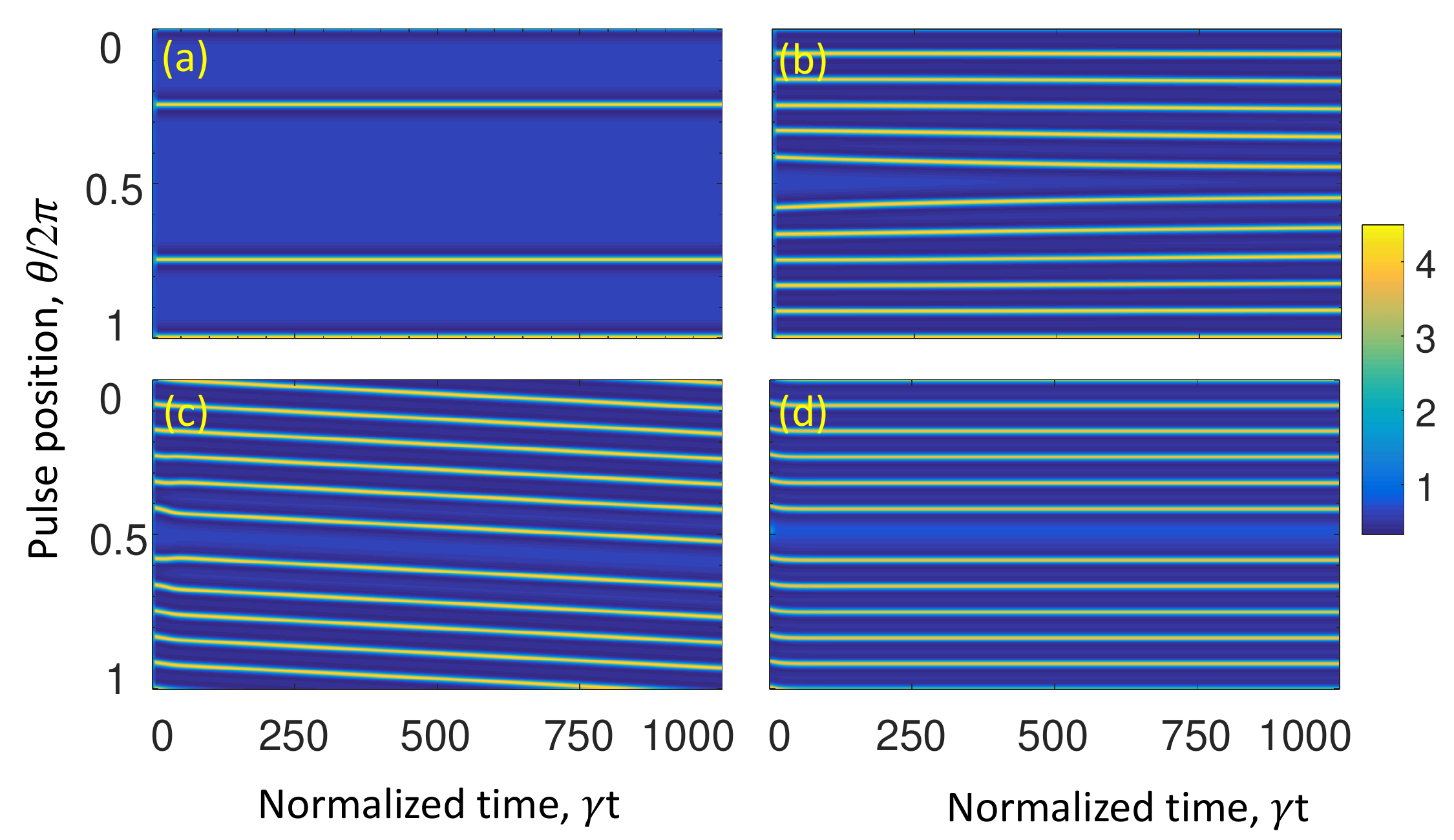}
\caption{ \small Multi-soliton microcombs in a monochromatically-pumped resonator (a, b), a microresonator with mode anti-crossing (c), and a dichromatically-pumped resonator (d). Soliton peaks are illustrated using color density. In a microresonator with ideal second-order dispersion characterized by $D_\mathrm{int}(j-j_0)/\gamma=0.125(j - j_0)^2$ and pumped by a monochromatic laser ($F_0 = 1.3620$, $\sigma_0 = - 2\gamma$), when the intra-cavity solitons are far from each other (a), they can have arbitrary relative positions. When, on the other hand, they are closely packed (b), their interactions with each other and the CW background results in the even redistribution of the pulses around the resonator circumference. If the resonator modes hosting the micrcomb feature an avoided mode crossing (c), so that $D_\mathrm{int}(j-j_0)/\gamma=0.125(j - j_0)^2 + (2.5)/(j - j_0 - 48.5)$, the pulse positions, unlike (a), will not be arbitrary, nor will the solitons redistribute as they did in (b). A similar pulse trapping effect is observed when the resonator used in (a) and (b) is pumped by a second pump ($j_1 - j_0 = 12$) with intensity $F_1 = F_0 / 10$ which is detuned from its nearest resonance by $\sigma_1 = -2\gamma$. Total integration time is 2000 cavity photon lifetimes.
} \label{fig:Solint}
\end{figure}

The trapping of solitons at lattice sites defined by the beatnote between the pump and the dispersive wave, which was discussed in the previous Section, essentially shows that the frequency comb is locked to the frequencies of the dispersive wave and the pump and therefore the comb repetition rate is fixed. Mode locking of a multi-soliton frequency comb in a monochromatically pumped resonator does not impose requirements on the spacing between the solitons (so long as they are not too close to each other to interact). For few-soliton combs, the interaction of the solitons with each other and the CW background is weak and the relative pulse position is arbitrary, Fig.~\ref{fig:Solint}(a), whereas for an increased number of solitons closely packed in a resonator the mutual interaction of the pulses and the CW background tends to redistribute pulses with arbitrary separation evenly around the resonator circumference, Fig.~\ref{fig:Solint}(b). The presence of a DW peak in the microcomb power spectrum, e.g., one due to a mode anti-crossing, can avoid the redistribution of the solitons around the resonator into an equally-spaced grid, as we have shown in Fig.~\ref{fig:Solint}(c). The same effect is observed when a second pump with suitable frequency is added to drive the microresonator, Fig.~\ref{fig:Solint}(d). In what follows we numerically study soliton train ordering and show that the interaction of solitons with the modulated background (resulting from the beating of the pump and a DW or a second pump) causes crystallization of the soliton train.

As we discussed in Section \ref{sec:Xtal_DW}, the train of optical pulses corresponding to the mode-locked Kerr frequency comb can be synchronized with the beatnote between the DW, generated because of the specific frequency-dependent GVD of the resonator, and the pump light. The DW and the frequency comb are generated simultaneously, so it is hard to study their synchronization mechanism. In this Section, we consider a different situation and introduce a second CW pump harmonic to a resonator characterized by an ideal quadratic anomalous GVD. We provide numerical simulation results which support the idea rendered in Fig.~\ref{fig:BCScheme}, i.e., locking the second pump (Pump 2) to the comb supported by the main pump (Pump 1) through tuning the frequency of Pump 2. Pump 2 resembles a DW, however it can be added after the Kerr frequency comb is formed. We see that the Kerr comb generated initially corresponds to a train of dissipative solitons having arbitrary relative positions. Switching the second pump on results in shifting the pulses to the positions defined by the beatnote between Pumps 1 and 2, and in the synchronization of the frequency comb with the beatnote signal of the pump waves. In this case, the phase locking of the microcomb to the second pump can be explained by the dynamic interaction of the pulse with the CW background modulation since the time scale of the modification of the frequency comb can be longer compared to that of establishing steady state for the second pump in the resonator. The locking also results in modified repetition rate of the frequency comb, similar to the cases studied in the previous Section.

\subsection{Locking range}

\begin{figure}[tbp]
  \centering
  \includegraphics[width=8.5cm]{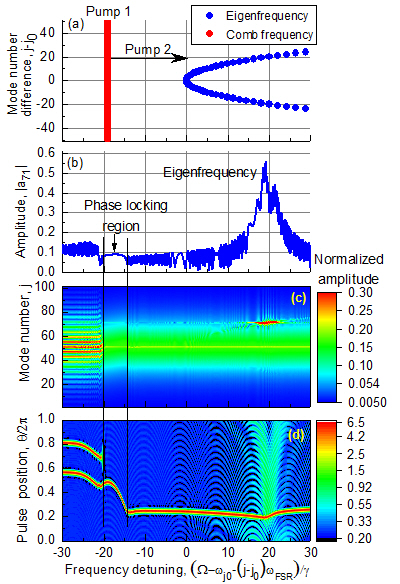}
\caption{ \small Modification of the comb parameters when the frequency of the second pump is scanned through the corresponding mode. The total number of modes in the model is 101, and the mode labeled 51 is pumped by the Pump 1. The relative frequency positions of the cold resonator modes and the frequency comb harmonics generated with Pump 1 having detuning $\sigma_0 = -19 \gamma$ are shown in panel (a). The resonator is characterized with the GVD parameter $D_2=0.1\gamma$ (anomalous GVD), so that the residual frequency is $D_\mathrm{int}(j - 51)/\gamma=0.05(j - 51)^2$. Pump 2 frequency detuning $20 \gamma$ corresponds to the position of the harmonic $a_{71}$ for the case of absence of the second pump.  The pulse amplitude $|A|$ is shown by the color density.   The main pump wave is characterized with the normalized power $|F_0|^2=20$. The Pump 2 amplitude is defined as $|F_1|=|F_0|/2$. The absolute amplitude of the harmonic $A_{71}$ pumped with the second pump is shown in panel (b). Corresponding frequency spectrum and the pulse position in the retarded frame of reference are shown in panel (d).
} \label{fig:fscan}
\end{figure}

To understand the effect of Pump 2 on the Kerr comb generation, we performed a numerical experiment in which we adiabatically scanned the frequency of this pump through the corresponding mode of the resonator and observed the dynamic behavior of the system. The results of the simulations are summarized in Fig.~\ref{fig:fscan}.

The frequency of Pump 1 was red-detuned with respect to the corresponding mode and was fixed. The power of the first pump was selected in a way to excite a mode-locked Kerr frequency comb narrow enough to guarantee avoiding the influence of the finite number of modes considered. The Kerr comb was excited in the system by introducing finite energy to the comb modes at the beginning of the simulation (to enable hard excitation of the comb). The second pump was tuned through the corresponding mode, as shown in Fig.~\ref{fig:fscan}(a). The time scale of the frequency tuning exceeded 100 ringdown times to ensure the adiabaticity of the process. In this and the following figures, the detuning of the pump frequency $\Omega_2$ from the corresponding mode is expressed in the dimensionless units $[\Omega_2-\omega_{j_0}-D_1(j-j_0)]/\gamma$, where as before $D_1 = \omega_{\mathrm{FSR}}(j_0)$ is the FSR of the resonator at the frequency of Pump 1, and $\gamma$ is the HWHM of the cavity resonance. The detuning defined in this way for the first pump corresponds to $(\Omega_1-\omega_{j_0})/\gamma = \sigma_0/\gamma$. The mode-locked Kerr frequency comb is generated for the red-detuned pump, so that $\sigma_0 < 0$.

Figure~\ref{fig:fscan} illustrates the following processes. An optical frequency comb comprising two solitons confined in the resonator is at first generated from random initial conditions. Pump 2 is initially detuned far to the red from its neighboring resonance. As it is scanned through and at a particular detuning in the vicinity of the corresponding frequency comb harmonic, the entire comb locks to Pump 2. To visualize the locking, we plot the absolute value of the amplitude of the light localized in the pumped mode, Fig.~\ref{fig:fscan}(b). The intensity of the power in this pumped mode does not change much when locking occurs. Interestingly, the locking of the comb annihilates one of the intra-cavity pulses and creates a comb with a smooth power spectrum envelope with single-FSR spacing between its adjacent teeth, as noted in the locking region marked on Fig.~\ref{fig:fscan}(c). It is possible to see in Fig.~\ref{fig:fscan}(c), that the center of the comb slightly shifts while the frequency of Pump 2 is tuned. Continuing the frequency scan of Pump 2, we also note that, unlike before the locking condition prevailed, the remaining pulse follows the moving pattern of the modulated waveform pedestal, Fig.~\ref{fig:fscan}(d), in agreement with the predictions of the theoretical framework presented in Section \ref{sec:formulation}. In contrast to the constant slope of the motion of the pulse peak noted in Fig.~\ref{fig:DWXtal}, the soliton peak follows a curved trajectory in Fig.~\ref{fig:fscan}(d). The behavior of the microcomb in the locking region resembles the expansion of accordion bellows and roots in the continuous change of the microcomb repetition rate as the second pump is tuned into resonance. Further scanning of the frequency of Pump 2 unlocks the comb, marked by the independent motion of the cavity soliton on top of the modulated intra-cavity field background. When the frequency of Pump 2 coincides with the peak of its cold neighboring resonance, it couples maximum power into the resonator, leading to a significant increase in the power of the relevant comb harmonic, Fig.~\ref{fig:fscan}(b). Even though the intracity power is maximazed for the second pump at this stage, the comb is not locked to it. The trajectory of the intracavity pulse changes only slightly in this region. The general stability of the comb degrades because of the large intra-cavity power coupled in by the Pump 2.

It is noteworthy that the coexistence of the phase-locking detuning region (between the frequency comb and Pump 2) and the resonant detuning region (corresponding to the position of the resonator ``cold'' cavity mode) expands the notion of Feshbach resonance in microcombs. It was recently shown that Feshbach resonances can occur in microresonator-based frequency combs when higher-order dispersion or other perturbations such as nonlinear absorption, gain, loss, or frequency filtering in the resonator tends to couple the intra-cavity soliton and the CW background, and results in modification of the soliton parameters \cite{matsko2015noise, matsko2015feshbach}. It was shown theoretically \cite{matsko2015feshbach}, and verified experimentally \cite{guo2016universal}, that adding a small modulation sideband to the pump light allows measuring the position of the cold resonance, and hence, inferring the detuning of the pump from the position of the cold cavity mode. We have shown here that a similar experiment can be performed using Pump 2 which enables the observation of comb harmonics as well as the spectral position of the corresponding cold cavity mode.

\subsection{Tuning microcomb repetition rate}

\begin{figure}[tbp]
  \centering
  \includegraphics[width=8.5cm]{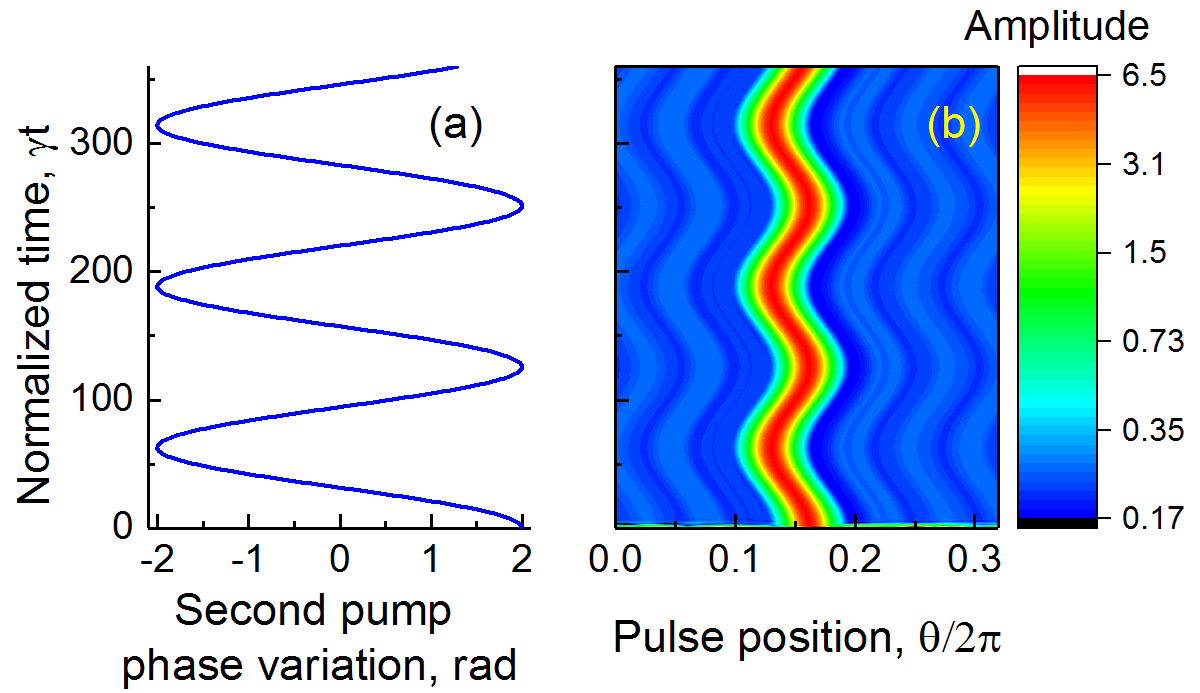}
\caption{ \small Behavior of the pulse position, (b), when the phase of the second pump is adiabatically modulated as shown in panel (a). The pulse position changes synchronously with the phase change. The zero frequency of the second pump corresponds to the position of the harmonic $a_{71}$ for the case of absence of the second pump.  The pulse amplitude $|A|$ is shown by the color density.  The main pump wave is characterized with the normalized power $|F_0|^2=20$ and frequency detuning $\sigma_0=-19 \gamma$. The second pump is defined as $|F_1|=|F_0|/5$ and is red detuned from the corresponding "cold" mode by $39\gamma$ and phase modulated with rate $\exp [2\imi\cos(0.05\gamma t)]$. The resonator is characterized with the GVD parameter $D_2=0.1\gamma$ (anomalous GVD), so that the residual frequency is $D_\mathrm{int}(j - 51)/\gamma=0.05(j - 51)^2$.
} \label{fig:wiggling}
\end{figure}

The trapping of cavity solitons by the optical lattice created as a result of the beating between Pump 1 and Pump 2 in the locking region is observed clearly in Fig.~\ref{fig:wiggling}, where the phase of the second pump is modulated adiabatically. The soliton peak, specified in red in the color density plot of Fig.~\ref{fig:wiggling}(b), follows the pattern of the modulated pedestal of the intra-cavity waveform as the phase of Pump 2 is altered. The soliton behaves as a trapped particle here, following the trap position.

Pulse position in Fig.~\ref{fig:wiggling} is shown in the co-rotating frame of reference. When the soliton repetition rate coincide with the FSR of the resonator at the frequency of the optical pump ($D_1$), the pulse does not move in this reference frame. Modification of the pulse peak position results from the deviation of the repetition rate from the FSR. Therefore, phase modulation of the second pump results in the modulation of the microcomb repetition rate. Below, we explore this phenomenon in further detail.

\subsection{Comb locking dynamics}

\begin{figure}[tbp]
  \centering
  \includegraphics[width=8.5cm]{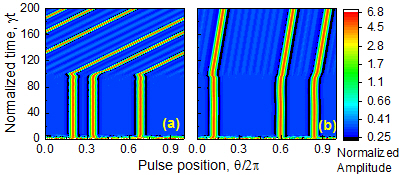}
\caption{ \small Formation of the optical crystal consisting of three solitons for two different frequency positions of the dichromatic optical pump. The pulse amplitude $\abs{A}$ is shown by the color density. The total number of modes in the model is 101. The main pump wave is applied to the mode labeled $j_0=51$, while the second pump is applied to mode $j_1=61$ [$j_1=71$] in panel (a) [(b)]. The main pump wave is characterized with the normalized power $|F_0|^2=20$ and frequency detuning $\sigma_0 = -14 \gamma$. The second pump is defined as $|F_1|=(|F_0|/5)[1/2 + (1/\pi) \mathrm{arctg}((\gamma t-100)/3)]$ and is detuned from the corresponding mode by $-18.4 \gamma$ ($-34.1 \gamma$ for the case shown in panel (b)). The values of the detunings can be understood from Fig.~\ref{fig:fscan}(a). To achieve the locking the second pump has to be in the vicinity of the unperturb comb harmonics. The resonator is characterized with GVD parameter $D_2=0.1\gamma$ (anomalous GVD), so that the residual frequency is $D_\mathrm{int}(j-51)/\gamma=0.05(j - 51)^2$.
} \label{fig:Xtalform}
\end{figure}

To explore the locking dynamics of the frequency comb to the second pump we consider the case of the adiabatic switching of the pump. In Fig.~\ref{fig:Xtalform}, we show two examples, where the second pump is applied coherently at $\gamma t = 100$ to two different cavity resonances separated by 10 FSRs. Consistent with Eq.~(\ref{eq:bichrom:SSpositions}), a lattice is created which traps the intra-cavity solitons and carries them around the resonator along with the modulated background. The pulses in both examples are formed starting from random initial conditions. Since the resonator is characterized by second-order dispersion and all higher-order dispersion terms are negligible, the generated cavity solitons are stationary in the rotating reference frame defined by Eq.~(\ref{eq:bichrom:rotframe}) before the addition of the second pump. However, when the second pump is tuned in, it changes the repetition rate of the microcomb and as a result the pulses start to move around trapped in the beatnote fringes. Two different excitation modes are considered for Pump 2 to illustrate the change of the beatnote period and to show that the pulse locks to the beat note pattern. Comparison of Figs.~\ref{fig:Xtalform}(a) and (b) also shows that the change of repetition rate depends on the power and frequency (spectral position) of the second pump. The two different frequency detunings in the two panels of Figs.~\ref{fig:Xtalform} have led to different modifications of the group velocities of the pulses. Dynamics of the frequency comb locking to the second pump depicted in Fig.~\ref{fig:Xtalform} emphasizes that the locking mechanism can be considered from the perspective of locking the pulses to the beatnote signal. This spatiotemporal picture is equivalent to the frequency domain perspective based on phase-locking of the microcomb to another oscillator.

\begin{figure}[tbp]
  \centering
  \includegraphics[width=8.5cm]{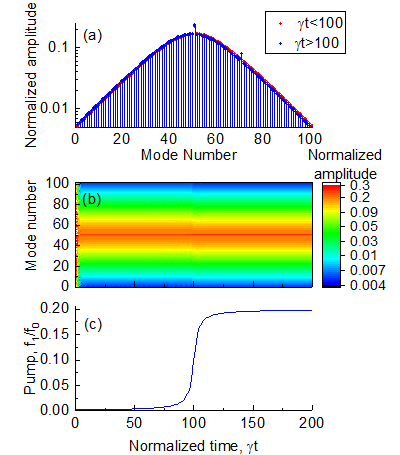}
\caption{ \small Change of the steady state power spectrum envelope of the Kerr comb (a), and the temporal evolution of this envelope (b), for the second pump with power temporal profile shown in panel (c). Parameter values used in the numerical simulation are similar to those of Fig.~\ref{fig:Xtalform}(b).
} \label{fig:combchange}
\end{figure}

Figure \ref{fig:combchange} illustrates the change in the comb power spectrum when the power of the second pump is increased adiabatically such that it adds coherently to the spectrum of a soliton created by the first pump (the red power spectrum in panel (a)). It is observed that after $\gamma t = 100$, when the power of the second pump becomes significant compared to the corresponding harmonic of the frequency comb, the entire power spectrum is slightly altered so that its peak (signifying the soliton spectral center) is shifted away from the strong comb harmonic excited by the second pump. The shift corresponds to a change in the soliton repetition rate. Parameters used in the numerical simulation are similar to those of Fig.~\ref{fig:Xtalform}(b). 

\begin{figure}[tbp]
  \centering
  \includegraphics[width=8.2cm]{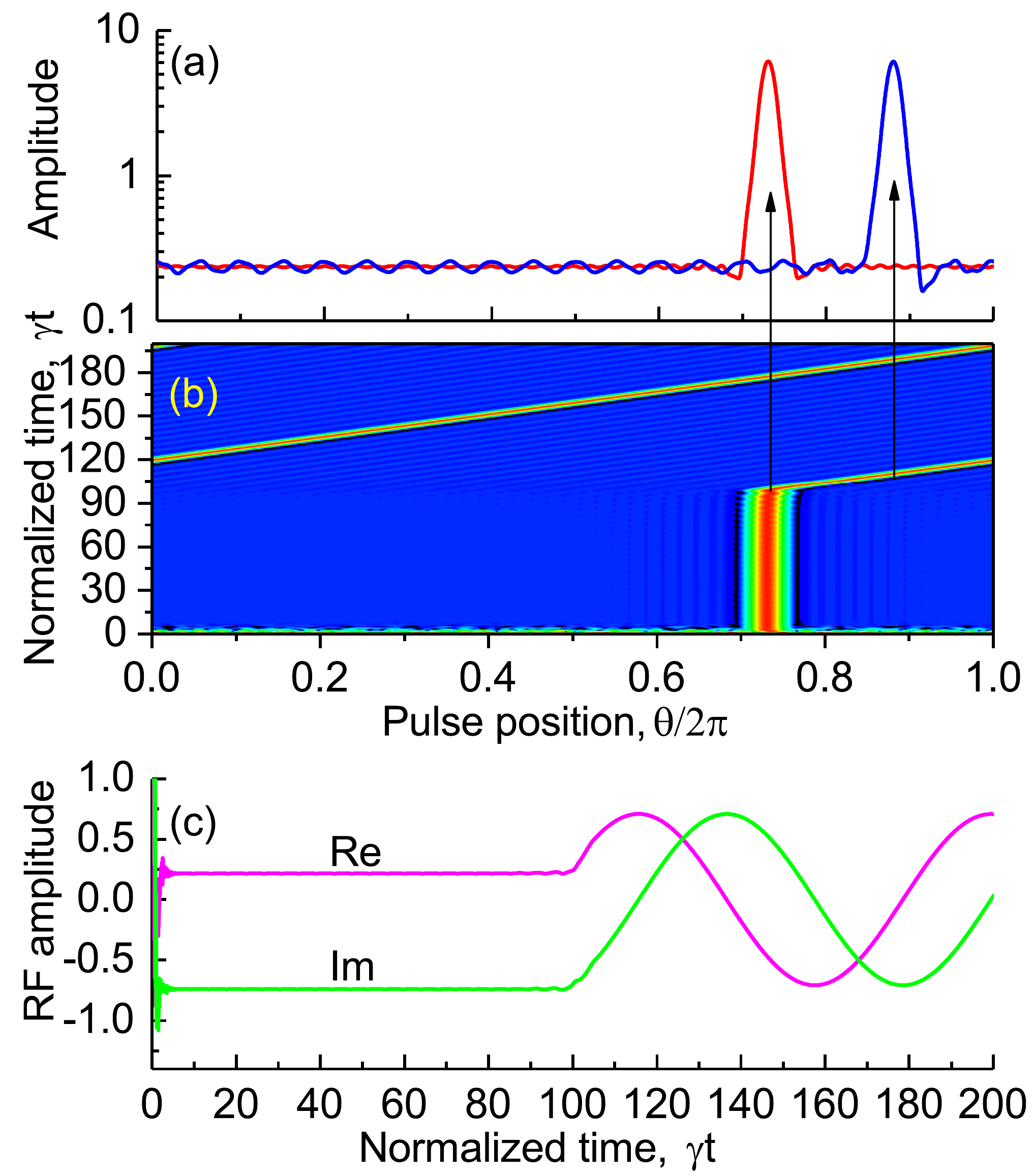}
\caption{ \small Temporal pulse modification (panels (a) and (b)) for the frequency comb shown in Fig.~(\ref{fig:combchange}). Panel (c) illustrates change of the radio frequency (RF) signal created by central harmonics of the frequency comb on a fast photodiode. We take into consideration only harmonics $a_{35}$ through $a_{67}$, while the first pump is at mode $j_0=51$ and the second pump at mode $j=71$.
} \label{fig:drift}
\end{figure}

The change induced in the power spectrum of the generated microcomb of Fig.~\ref{fig:combchange} is also noted in its spatiotemporal waveform, Fig.~\ref{fig:drift}. The pulse shape does not change much and the background modulation remains relatively small, Fig.~\ref{fig:drift}(a), but the frequency of the RF signal, Fig.~\ref{fig:drift}(c)) changes in a way akin to Fig.~\ref{fig:DW}(d)). Comparing these figures, one can conclude that the second pump impacts the frequency comb similarly to a dispersion wave \cite{taheri2016highorderdisp}: the center of the comb envelope shifts and the repetition rate of the soliton changes.

\begin{figure}[tbp]
  \centering
  \includegraphics[width=8.5cm]{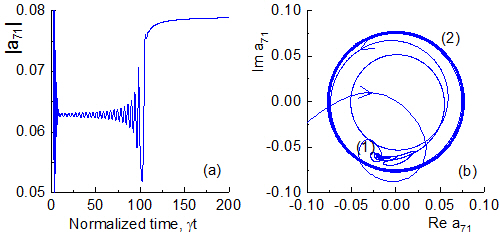}
\caption{ \small (a) Temporal behavior of the intensity of the microcomb harmonic driven by the second pump. When the power of Pump 2 is small ($\gamma t \approx 0$, see panel (c) of Fig.~\ref{fig:combchange}), the intensity of the harmonic reaches its steady state (attractor (1) in panel (b)) corresponding to the generation of a frequency comb by the main pump. Increase of the power of the second pump increases the power of the comb harmonic and oscillations occur owing to the beating of this comb tooth and Pump 2. As the pump power exceeds a certain value, it locks to the comb harmonic (and the entire comb), the beating stops, and the intensity of the harmonic reaches its steady-state value ($\gamma t > 100$, cf., panel (c) of Fig.~\ref{fig:combchange}). Since the repetition frequency of the microcomb changes when it becomes phase locked to the second pump, attractor (2) corresponding to this steady state is represented by a circle in the phase diagram, as depicted in panel (b).
} \label{fig:attractors}
\end{figure}

In Fig.~\ref{fig:attractors}(a), we observe the temporal variations of the intensity of the microcomb harmonic in the vicinity of Pump 2. The intensity of this comb tooth stabilizes within a few ringdown times after switching the first pump on. When the power of Pump 2 is small ($\gamma t < 100$ in Fig.~\ref{fig:combchange}(c)), slow beating is noticed between the comb harmonic and Pump 2. Increase of Pump 2 power above a certain threshold results in the locking of the neighboring comb harmonic (and the entire comb) to Pump 2. As a consequence, no beating is observed in the intensity of this comb tooth. Moreover, the power of the comb harmonic increases, signifying that, after locking, the energy of Pump 2 is being consumed by the comb.

The phase diagram of Fig.~\ref{fig:attractors}(b) shows that upon switching the second pump on, the system moves from one attractor (corresponding to the unperturbed frequency comb generated by Pump 1) to another (related to the comb state defined by both pumps). While the first attractor is a point in the phase diagram, the second attractor appears as a circle because the frequency of the comb harmonic neighboring Pump 2 shifts from its steady state position when locking kicks in.

The data presented above confirms that the frequency comb generated by the main pump can become phase locked to the second pump. This property is useful for comb stabilization. Usage of two coherent lasers with known frequencies for microcomb formation may result in the generation of Kerr frequency combs with well-defined repetition rate and offset frequency. No electronic locking is needed to support the generation of such frequency combs. It is essential, though, to introduce the second pump in a proper way. A dichromatic pump applied to the cold resonator modes may generate, through four-wave mixing, a spectrum containing merely a few harmonics. A broadband comb can emerge if the conditions of hard excitation are sustained and the second pump has proper power to support the phase locking mechanism.

\subsection{Breathers due to the second pump}

\begin{figure}[tbp]
  \centering
  \includegraphics[width=8.5cm]{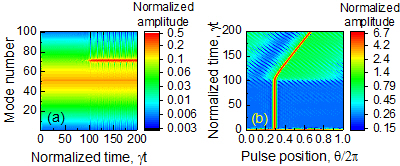}
\caption{ \small Breathing behavior of a dually-pump microresonator. Frequency comb power spectrum (a), and temporal evolution of the intra-cavity waveform (b), for the main pump wave applied to mode 51 and the second pump applied to mode 71 of the 101 modes considered in the numerical integration. The main pump wave is characterized by the normalized power $|F_0|^2=20$ and frequency detuning $\sigma_0 = -19 \gamma$. The second pump is defined as $|F_1|=(|F_0|/4)[1/2 + (1/\pi) \mathrm{arctg}((\gamma t-100)/3)]$ and is detuned from the corresponding mode by $2.5 \gamma$. The resonator is characterized with GVD parameter $D_2=0.1 \gamma$. 
} \label{fig:feshbach}
\end{figure}

Tuning the second pump in the vicinity of the corresponding cold cavity mode couples a significant amount of power to the resonator, as illustrated by Fig.~\ref{fig:fscan}(b). This increase does not lead to phase locking of the soliton, but creates an asymmetric breather, Fig.~\ref{fig:feshbach}(a). The breather has a peculiar power spectrum, which is more pronounced in the spectral part of the frequency comb in the neighborhood of the second pump. The pulse corresponding to this breather does not demonstrate significant changes in its duration, Fig.~\ref{fig:feshbach}(b). The intra-cavity pulse and the beatnote signal of the pumps are not locked. However, the repetition rate of the frequency comb is influenced by the strong intra-cavity power, as suggested by panel (b) of Fig.~\ref{fig:feshbach}.

In general, phase locking of the frequency comb to the second pump is observed when Pump 2 has significantly lower power compared with the first pumps. In the case of comparable pump powers, the system becomes unstable due to the competition between the frequency combs excited by each of the pumps. Tuning the second pump to the cold resonance increases the instability even further.

\section{Conclusion}
\label{sec:conclusion}
We have studied both analytically and numerically the formation of coherent mode-locked Kerr frequency combs via dichromatic coherent pumping. We have shown that this pumping scheme can lead to forming frequency combs locked to the frequencies of both pumps, thereby stabilizing the microcomb and fixing its repetition rate. The phase locking reveals itself in the trapping of cavity solitons in the grid defined by the beatnote of the two pumps which modulates the CW background of the intra-cavity waveform. Our results are important for complete stabilization of Kerr frequency combs and explain the generation of soliton crystals in microresonators supporting dispersive wave emission.

\section{Authors contributions}
All authors contributed to conceiving the idea and participated in preparing the manuscript. HT developed the theory and performed the analytical calculations. HT and AM performed the numerical simulations.

\bibliographystyle{unsrt}
\bibliography{references}
%
%
%
%
%

\end{document}